\def\year{2025}\relax

\documentclass[letterpaper]{article} 
\usepackage{aaai22}  
\usepackage{times}  
\usepackage{helvet}  
\usepackage{courier}  
\usepackage[hyphens]{url}  
\usepackage{graphicx} 
\usepackage{natbib}  
\usepackage{caption} 
\DeclareCaptionStyle{ruled}{labelfont=normalfont,labelsep=colon,strut=off} 
\usepackage{algorithm}
\usepackage{algorithmic}
\usepackage{tabularx}

\usepackage{array, makecell} 
\usepackage{multirow}

\usepackage{xcolor}

\usepackage{newfloat}
\usepackage{listings}
\floatstyle{ruled}
\newfloat{listing}{tb}{lst}{}
\floatname{listing}{Listing}
\usepackage{booktabs}
\usepackage{amssymb} 
\usepackage{graphicx}
\usepackage{subcaption}
\usepackage{url}

\newcommand{\descr}[1]{\smallskip\noindent\textbf{#1}}

\title{Exploring Left-Wing Extremism on the Decentralized Web: An Analysis of Lemmygrad.ml }

\author {
    Utkucan Balci \textsuperscript{\rm 1} 
    Michael Sirivianos \textsuperscript{\rm 2}
    Jeremy Blackburn \textsuperscript{\rm 1}
}
\affiliations {
    
    \textsuperscript{\rm 1} Binghamton University\\
    \textsuperscript{\rm 2} Cyprus University of Technology\\

    ubalci1@binghamton.edu, michael.sirivianos@cut.ac.cy, jblackbu@binghamton.edu
}

\usepackage{bibentry}

\begin{document}

\maketitle

\begin{abstract}
    This study\footnote{The original version of this paper published in DeWeb 2024. This is an updated version including only posts local to Lemmygrad.ml. Please cite accordingly.} investigates the presence of left-wing extremism on the Lemmygrad.ml instance of the decentralized social media platform Lemmy, from its launch in 2019 up to a month after the bans of the subreddits r/GenZedong and r/GenZhou.
    We conduct a temporal analysis on Lemmygrad.ml's user activity, with also measuring the degree of highly abusive or hateful content.
    Furthermore, we explore the content of their posts using a transformer-based topic modeling approach.
    Our findings reveal a substantial increase in user activity and toxicity levels following the migration of these subreddits to Lemmygrad.ml.
    We also identify posts that support authoritarian regimes, endorse the Russian invasion of Ukraine, and feature anti-Zionist and antisemitic content.
    Overall, our findings contribute to a more nuanced understanding of political extremism within decentralized social networks and emphasize the necessity of analyzing both ends of the political spectrum in research.  
   
  \end{abstract}
  
  \section{Introduction}
  \label{sec:intro}

  In the evolving context of social media, the decentralization of the Web has emerged as a pivotal transformation by creating additional space for online communities that advocate for a more democratic and user-empowered Internet.
  Although this paradigm shift can promote innovation and freedom of expression in many contexts, like many centralized social media platforms~\cite{grover2019detecting, perez2020trend, hine2017kek}, it has inadvertently become a safe space for extremist ideologies. 
  So far, the academic community has primarily focused on the proliferation of far-right extremism within these decentralized frameworks. 
  Particularly, attention has been directed towards specific platforms that have become notorious for their role in amplifying far-right ideologies.
  An example platform in this regard is Gab~\cite{zannettou2018gab,thiel2022gabufacturing}, which has been found to serve as a breeding ground for hate speech and extremist content.
  
  In contrast, the exploration of left-wing extremism on decentralized social media platforms remains overlooked so far. 
  This paper aims to bridge this gap by focusing on Lemmygrad.ml, a Lemmy instance which defines itself as a collection of Marxist communities~\cite{Lemmygradml}.
  By examining content and user interactions on Lemmygrad.ml from its inception in 2019 to one month after the migration of left-wing extremist (specifically, tankie) communities that were banned from Reddit, we aim to explore the extent of left-wing extremism on the Decentralized Web.
  
  In our analysis, we aim to contribute to the existing body of knowledge on the Decentralized Web, pointing out the double-edged sword it represents in terms of promoting democratic discourse on the one hand and enabling extremist content on the other. 
  In doing so, we call for a more nuanced understanding of the impacts of Web decentralization and the need for innovative governance mechanisms that safeguard against the proliferation of extremist ideologies while upholding the principles of freedom and democracy that are the basis of the Decentralized Web.

  \section{Background \& Related Work}
  \label{sec:background-related-work}

  The rise of online communities has fundamentally transformed the dynamics of political discourse by enabling the rapid spread of extremist ideologies. 
  Research  on online extremism has predominantly focused on centralized platforms (e.g., Twitter and Reddit), examining how extremists spread their ideology~\cite{perez2020trend, benigni2017online,zannettou2018origins, grover2019detecting}.
  Current work on extremism on these platforms mainly aim to understand the characteristics of far-right extremism online~\cite{scrivens2020measuring, zannettou2017web}, explore the role of social media in radicalization processes~\cite{ribeiro2020auditing}, and examine echo chamber effects~\cite{efstratiou2023non}.
  
  Parallel to the discourse on online extremism, the concept of the Decentralized Web has gained attention from researchers, with an emphasis on its potential to democratize social media platforms and mitigate the risks associated with centralized data control~\cite{kwet2020fixing,abbing2023decentralised,la2022information}. 
  Focusing on their peer-to-peer nature and community-driven governance models, research~\cite{cava2023drivers,struett2023can, bin2022toxicity, raman2019challenges} on Mastodon and the broader Fediverse have highlighted the architectural, behavioral, and operational distinctions that set these platforms apart from their centralized counterparts.
  While extensive research has been conducted on right-wing extremism within decentralized networks~\cite{zannettou2018gab,gerster2023hydra,thiel2022gabufacturing}, the exploration of far-left extremism on these platforms is still in its infancy.
  
  \descr{What is Lemmy?}
  Lemmy is an open-source, self-hosted Decentralized Web platform that provides instances for news aggregation and discussion forums~\cite{Lemmy}.
  These instances operate similar to centralized news aggregation platforms like Reddit, where users can create posts, comment on them, and cast votes.
  Additionally, Lemmy enables its users to communicate across different instances. 
  To date, there is limited literature on Lemmy.
  \cite{nunes2023user} analyzed sentiments on Lemmy to understand user migration from Reddit with analyzing nearly 50K posts without touching its extremist content. 
  Notably, this work find positive sentiments toward Lemmy and criticism of Reddit among the users of this platform.
  
  \descr{What is Lemmygrad.ml?}
  Lemmygrad.ml is a Marxist/Leninist Lemmy instance that mirrors Reddit and hosts various left-wing online communities.
  Similar to other Lemmy instances, users in this platform can create communities to post content and engage in discussions around them, by abiding the rules set by Lemmygrad.ml's administrators.
  To date, there has not been any research exploring Lemmygrad.ml.

  \section{Dataset}
  \label{sec:dataset}

\begin{figure}[ht!]
    \centering
    \includegraphics[width=1\columnwidth]{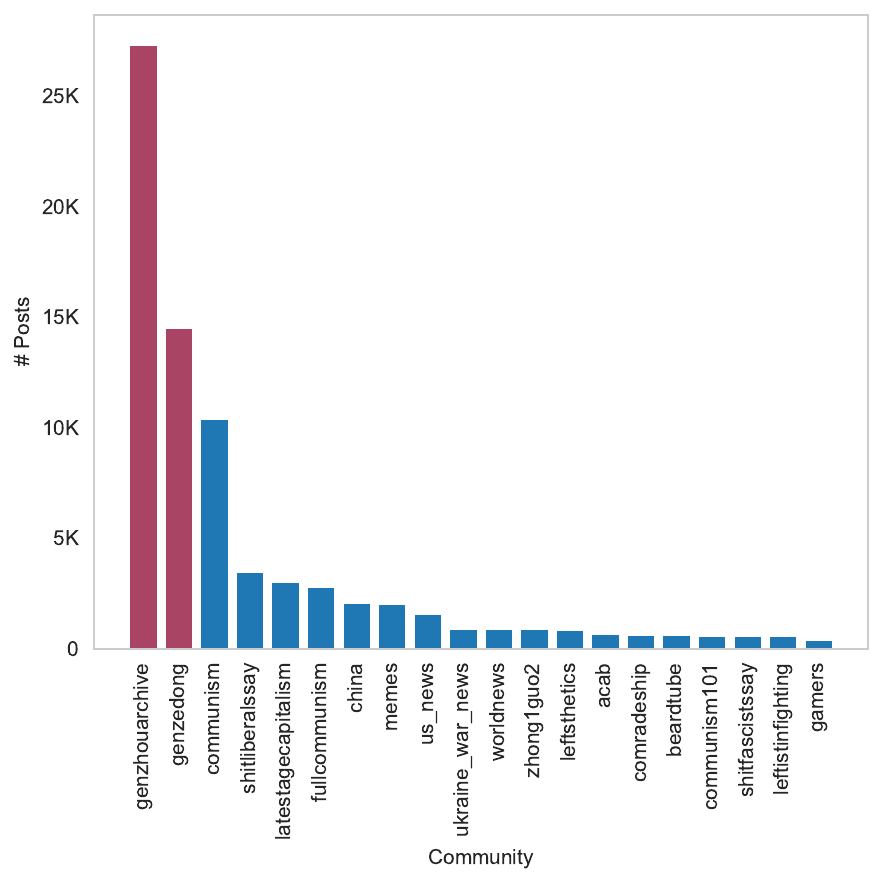}
    \caption{Top 20 communities of Lemmygrad.ml. Communities related to tankie subreddits are colored as red.}
    \label{fig:lemmy_top}
  \end{figure}

  Using a custom crawler, we collect 91,271 posts from 465 communities (Lemmy's equivalent of subreddits) between August 17, 2019, and April 30, 2022. 
  Figure~\ref{fig:lemmy_top} displays the top 20 communities of Lemmygrad.ml according to their total number of posts.
  The two most popular communities, c/genzhouarchive and c/genzedong, are related to left-wing extremist subreddits that have faced restrictions from Reddit, where r/GenZedong was quarantined for spreading misinformation~\cite{redditgenzedong}, and r/GenZhou was banned for ban evasion related to r/GenZedong's quarantine~\cite{redditgenzhou}. 
  Our manual inspection on the banners of these subreddits finds that they fit Petterson's description of contemporary tankies~\cite{petterson2020apostles}, where they declare their support for Actually Existing Socialist (AES) countries, while also identifying themselves as Marxist-Leninists. 
  The most popular community, c/genzhouarchive, is an archive of the subreddit, r/GenZhou. Moreover, the second most active community, c/genzedong, further solidifies Lemmygrad.ml's role as a Reddit alternative for tankies.

  \begin{figure}[ht!]
    \centering
    
    \begin{subfigure}{\columnwidth}
      \includegraphics[width=\linewidth]{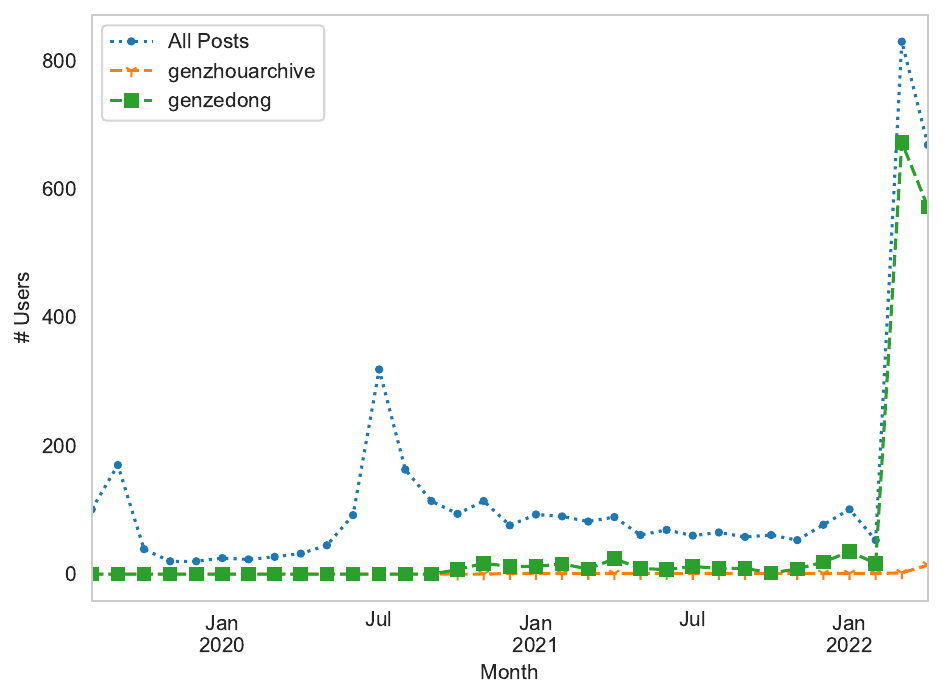}
      \caption{}
      \label{fig:lemmy_monthly_users}
    \end{subfigure}

    \vspace{1em}

    \begin{subfigure}{\columnwidth}
      \includegraphics[width=\linewidth]{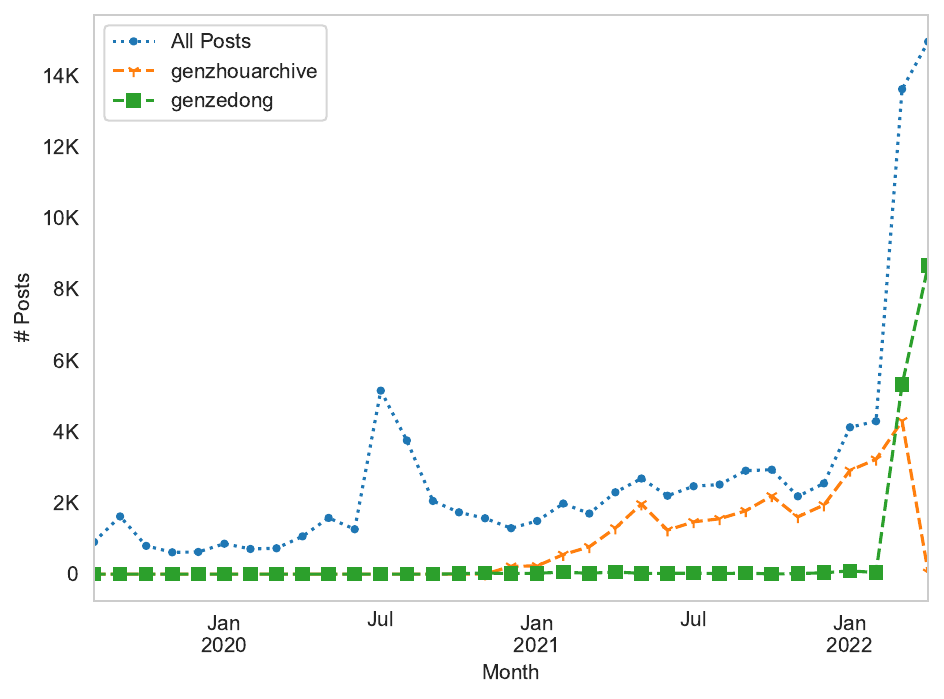}
      \caption{}
      \label{fig:lemmy_monthly}
    \end{subfigure}
  
    \caption{Monthly total number of (a) active users and (b) posts of c/genzhouarchive, c/genzedong, and all communities of Lemmygrad.ml. Note that genzhou archive drop off after the ban on r/GenZhou since it is acting as an archive for this subreddit.} 
  \end{figure}

  \section{Temporal Analysis}
  
  We analyze the temporal information of Lemmygrad.ml, focusing on their popular tankie communities. 
  We first examine monthly user activities to gain an understanding of the Lemmygrad.ml's growing popularity and the impact of quarantines on tankie subreddits.
  Subsequently, we explore the evolution of highly toxic posts to complement our findings from the analysis of monthly user activity.

  \descr{Monthly User Activity.}
  To investigate the impact of quarantines on the popularity of r/GenZedong and r/GenZhou on Lemmygrad.ml, we examine the monthly active user counts (Figure~\ref{fig:lemmy_monthly_users}) and the the monthly post counts (Figure~\ref{fig:lemmy_monthly}) for c/genzhouarchive, c/genzedong, and all of Lemmygrad.ml.
  We find that c/genzedong shows a sharp rise in monthly user activity after r/GenZedong's March 2022 quarantine, peaking at 8,666 posts from 572 users in April 2022.
  Before the month of quarantine, c/genzedong's highest user activity was in January 2022, with a maximum of 92 posts from 35 users.
  Conversely, as it is an archive community, the ban of r/GenZhou in April 2022 resulted in a heavy decline in c/genzhouarchive's user activity.
  The total number of posts dropped from 4,305 to 47, even though the count of active users peaked, rising from 2 to 14 monthly active users in the same period.
  We also notice a general uptick in Lemmygrad.ml's monthly posts, especially post-bans/quarantines, with the numbers rising from 4,297 posts by 53 users in February 2022 to 14,962 posts by 669 users in April 2022.
  Our findings suggest that, similar to their right-wing counterparts on Gab~\cite{thiel2022gabufacturing}, platform restrictions greatly increased the popularity of Lemmygrad.ml.
  
  \begin{figure}[ht]
    \centering
    \includegraphics[width=1\columnwidth]{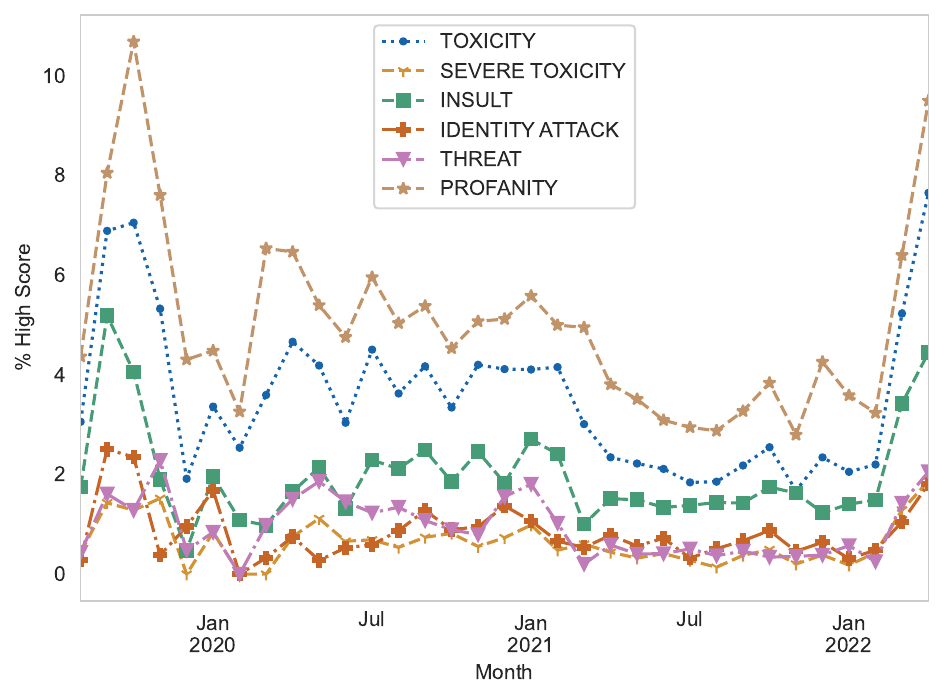}
    \caption{Monthly proportions of high scores of Lemmygrad.ml regarding Perspective API attributes.}
    \label{fig:lemmy_perspective}
  \end{figure}
  
  \descr{Monthly Perspective Analysis.}
  Perspective API~\cite{perspective} is a tool developed by Google which measures abusive or hateful content within texts, providing scores from 0 to 1, lowest to highest.
  Using Perspective API, previous research on Mastodon~\cite{he2023flocking} and Pleroma~\cite{hassan2021exploring} found the existence of highly toxic posts within these networks.
  Ribeiro et al.~\cite{horta2021platform} observed an increase in toxicity levels following the migration of users from the subreddit r/The\_Donald to the platform thedonald.win, using Perspective API scores as a metric for measurement.
  Building on this, we measure the monthly high Perspective API score proportions of the posts on Lemmygrad.ml.
  Following previous work~\cite{he2023flocking,hoseini2023globalization,balci2023beyond}, we consider posts $\geq 0.8$ as having high Perspective API scores.
  We use the production models~\cite{Jigsaw} of Perspective API, which have undergone testing processes in various domains and have been developed based on large volumes of comments annotated by humans.

  Figure~\ref{fig:lemmy_perspective} illustrates the monthly proportions of high scores from Perspective API models for c/genzhouarchive, c/genzedong, and all of Lemmygrad.ml.
  We find that in 2019, the year Lemmygrad.ml was founded, there was a notable peak in the frequency of high scores across 
  \texttt{TOXICITY}, \texttt{SEVERE\_TOXICITY}, \texttt{INSULT}, \texttt{IDENTITY\_ATTACK}, \texttt{THREAT}, and \texttt{PROFANITY} models.
  This peak occurred despite 2019's relatively low monthly activity, which averaged 911.6 posts.
  However, the subsequent years, 2020 and 2021, saw a substantial reduction in the average high scores for all models, cutting them by more than half.

  Interestingly, in March and April 2022, despite recording the highest number of monthly posts, the proportion of posts with high Perspective API scores reached new highs for toxicity and severe toxicity, and rose sharply for the other models.

  \descr{Takeaways.}
  We find left-wing extremists have substantial activity on Lemmygrad.ml. 
  Our analysis shows that Lemmygrad.ml's top two communities, c/genzedong and c/genzhouarchive, are associated with tankie subreddits, r/GenZedong and r/GenZhou.
  Furthermore, our findings suggest that tankies' platform migration resulted in an increase in user activity and toxicity on Lemmygrad.ml, highlighting the importance of considering the impact of platform migration to decentralized networks on online communities.

  \section{Topic Analysis}
  
  We aim to understand the content discussed in this platform by using a transformer-based topic modeling technique and explore their popular discussions to gain an understanding of Lemmygrad.ml.
  
  \descr{Topic model.}
  We leverage BERTopic~\cite{grootendorst2022bertopic}, which uses transformer-based sentence embeddings to generate embeddings of documents and has been previously used in studies to explore content of online communities~\cite{hanley2023happenstance,mekacher2023systemic,tahmasbi2025going}.
  This method automatically finds dense clusters of documents by first applying dimensionality reduction to the embeddings with UMAP~\cite{mcinnes2018umap}, and clustering them using HDBSCAN~\cite{mcinnes2017hdbscan}.
  Before training our model, we remove hyperlinks and user names from posts, and exclude deleted or removed posts. 
  Additionally, we observe that posts from the c/genzhouarchive contain the text: ``originally from r/GenZhou,'' so we also remove these references from the posts.
  To refine our analysis, we also manually inspect 100 random posts from each topic.

  \begin{table}[t]
    \centering
    \small
    \begin{tabular}{rlr}
    \toprule
    No. & Top 3 Keywords & Count \\
    \midrule
    1 & ridic, neatza, totes & 2,830                 \\ 
    2 & xd, tysm, ja & 1,181               \\  
    3 & dprk, korea, kim & 838             \\     
    4 & fascism, fascist, fascists & 729             \\     
    5 & comrade, thank, thanks & 661                \\  
    6 & communism, communist, my & 599           \\       
    7 & ukraine, nato, ukrainian & 478                 \\ 
    8 & gorb, bonk, eussr & 466           \\      
    9 & hes, he, him & 444                         \\ 
    10 & lemmy, lemmygrad, lemmygradml & 383                                \\
    11 & china, economy, socialism & 367                   \\              
    12 & nazis, nazi, hitler & 351                                \\ 
    13 & america,american,americans & 325                    \\              
    14 & mao, maoists, maoism & 322                     \\            
    15 & elaborate, why, point & 315                      \\          
    16 & cuba, cuban, castro & 291                \\                
    17 & nother, pic, saw & 279                                \\
    18 & star, watch, season & 261                                \\
    19 & israel, jews, jewish & 243                              \\  
    20 & genzedong, rgenzedong, banned & 238                          \\       
    \bottomrule
    \end{tabular}
    \caption{Top 20 topics discussed on Lemmygrad.ml.}
    \label{table:top_topics}
    \end{table}

  \descr{Popular discussion themes of Lemmygrad.ml.}
  Table~\ref{table:top_topics} presents the top 20 topics with most posts on Lemmygrad.ml. 
  We find that users on Lemmygrad.ml frequently discuss two AES countries, China and North Korea, with many posts expressing support for them. 
  An example post from c/genzedong:
  
  \begin{quote}
  
    \textit{``DPRK IS THE **ONLY** KOREA, IMPERIALISTS CONTINUE TO OCCUPY THE SOUTHERN REGION OF DPRK.''}
    \end{quote}
  
  In the topic related to Ukraine, we find many posts supporting or justifying the Russian invasion of Ukraine.
  An example post from c/politics:

  \begin{quote}
  
    \textit{``No one's asking you to be ok with all the horrors of war. But if we don't understand what started it, we won't be able to finish it. Ukraine's ethnic cleansing of Russians, suppression of the LPR and DPR, and flirtation with NATO must stop if ever this war will end.''}
    \end{quote}

  We find that discussions on the Israel-Palestine conflict primarily criticize Israel. 
  While many posts condemn antisemitism, we also encounter numerous posts that extend beyond criticizing Israel, displaying anti-Zionism and even antisemitism.
  An example post from c/communism:

  \begin{quote}
  
    \textit{``1. Jews arent a nation 2. Isael is a fake country.  3. The only thing which could be more fake than Isael and the ``jewish'' national identity would be for the fans of start trek in to create a language of their own similar to english, then enter the bourgeoisie and amass wealth and power, and then invade some imperialized nation 200 years later, settle in it and claim their own stark-treakish fake nation state. And they will justifyng this by quoting some science fiction book written in the 1900s which tells how are they the chosen people to colonize this new land.  Everyone identifing as a jew is a zionist and an enemy of the communist and anti imperialist movement.''}
    \end{quote}

  Besides relatively mundane topics, we see topics related to ideologies, economics, and countries, where posts predominantly reflect Marxist/Leninist or Maoist perspectives.

  \descr{Takeaways.}
  Our results show that users of Lemmygrad.ml frequently share posts that support authoritarian regimes, as seen in their support for China, North Korea, and Russia. 
  Moreover, their support can extend beyond backing these authoritarian regimes, even cheering on their violent actions, as evidenced by their posts on the Russian invasion of Ukraine.
  Additionally, we observe anti-Zionist and antisemitic behaviors, which show similarities to right-wing extremism~\cite{zannettou2020quantitative}.
  Our analysis suggests a concerning endorsement of authoritarian actions and extremist rhetoric on Lemmygrad.ml, further indicating that left-wing extremist communities on decentralized platforms should receive more attention from the academic community.

  \section{Conclusion}
  In this paper we explored a left-wing Lemmy instance, Lemmygrad.ml, which also serves as a hub for left-wing extremist subreddits that faced restrictions from Reddit.
  We find an increase in user activity and toxicity levels on Lemmygrad.ml following the migration of r/GenZedong and r/GenZhou.
  Furthermore, our analysis of the content revealed posts supporting authoritarian regimes, endorsing the Russian invasion of Ukraine, and exhibiting anti-Zionist and antisemitic rhetoric.
  Our findings underscore the importance of studying left-wing extremism on decentralized platforms alongside right-wing extremism to gain a comprehensive understanding of the full spectrum of political extremism on the Decentralized Web. 
  Additionally, by investigating the decentralized nature of platforms like Lemmygrad.ml, this paper contributes to the broader discourse on the implications of the Decentralized Web technologies. 
  
  \section{Future Work}
  While this work highlights the presence of left-wing extremism on the Decentralized Web, it is limited to a single instance on a single platform.
  Our study emphasizes the critical need to explore the prevalence and dynamics of left-wing extremism within the Decentralized Web.
  We also encourage future work to further investigate the distinctions between left-wing and right-wing extremism on these platforms, as well as to compare the characteristics of left-wing extremists on decentralized platforms with those on centralized social media platforms.
  
  \section{Acknowledgments}

  This material is based upon work supported by the National Science
  Foundation under Grant No. IIS-2046590 and the European Union under the project MedDMO (Grant Agreement no. 101083756).


  \newpage

  \end{document}